\documentclass[aps,onecolumn,noshowpacs,centertags]{revtex4}
\usepackage{graphicx}
\usepackage{graphics}
\usepackage{amssymb}

\begin{document}

\title{Early formation of galaxies initiated by clusters \\
of primordial black holes}

\author{V.~I.~Dokuchaev}\email{dokuchaev@lngs.infn.it}
 \affiliation{Institute for Nuclear Research of the Russian Academy of
 Sciences, Moscow, Russia}
\author{Yu.~N.~Eroshenko}\email{erosh@inr.npd.ac.ru}
 \affiliation{Institute for Nuclear Research of the Russian Academy of
 Sciences, Moscow, Russia}
\author{S.~G.~Rubin}\email{sergeirubin@list.ru}
 \affiliation{Moscow State Engineering Physics Institute,
31 Kashirskoe Sh., Moscow 115409, Russia}

\date{\today}

\begin{abstract}
Model of supermassive black holes formation inside the clusters of
primordial black holes is developed. Namely, it is supposed, that
some mass fraction of the universe $\sim10^{-3}$ is composed of
the compact clusters of primordial (relic) black holes, produced
during phase transitions in the early universe. These clusters are
the centers of dark matter condensation. We model the formation of
protogalaxies with masses about $2\,10^{8}{\rm M_{\odot}}$ at the
redshift $z=15$. These induced protogalaxies contain central black
holes with mass $\sim10^5{\rm M_{\odot}}$ and look like dwarf
spheroidal galaxies with central density spike. The subsequent
merging of induced protogalaxies and ordinary dark matter haloes
corresponds to the standard hierarchical clustering scenario of
large-scale structure formation. The coalescence  of primordial
black holes results in formation of supermassive black holes in
the galactic centers. As a result, the observed correlation
between the masses of central black holes and velocity dispersion
in the galactic bulges is reproduced.
\end{abstract}

\maketitle

\section{Introduction}

The problem of galaxy formation with supermassive central black
hole (BH) becomes more and more intriguing and ambiguous in view
of discovery of distant quasars at redshifts $z>6$ in Sloan
Digital Sky Survey \cite{z6}. The maximum observed red-shift
$z=6.41$ belongs to the quasar with luminosity corresponding to
the accretion onto BH with the mass $3\,10^{9}{\rm M_{\odot}}$
\cite{will03}. Such an early formation of BHs with masses
$\sim10^{9}{\rm M_{\odot}}$ is a serious difficulty for the
standard astrophysical models of supermassive BH formation in
galaxies supposing a fast dynamical evolution of the central
stellar clusters in the galactic nuclei (see e.~.g.
\cite{spitzer,saslaw,lightshap78,rees84,dokrev} and references
therein), a gravitational collapse of supermassive stars and
massive gaseous disks in galactic centers (see e.~.g.
\cite{rees84,bintrem87,el2,haehreeesnar98}), the multiple
coalescences of stellar mass BHs in galaxies (see e.~.g.
\cite{rees92,MouTan02,GebRicHo02,kawaguchi}) with the subsequent
multiple merging of galactic nuclei in collisions of galaxies in
clusters (see e.~g.
\cite{ValVal89,cattaneo05,komossa03,smirnova06,springel05,escala05,masjedi06}).
All standard astrophysical (or galactic) scenarios of supermassive
BH origin predict a rather late time of supermassive BH formation
in the galactic nuclei. An other difficulty is that all these
astrophysical scenarios are realized only in strongly evolved
galactic nuclei. In view of these problems the cosmological
scenarios of massive primordial BHs formation become attractive
\cite{zeld67,carr75,DolgovSilk93,quin,Ru1,carr05}. In cosmological
scenarios the seeds of supermassive BHs are formed long before the
formation of galaxies. These primordial black holes (PBH) can be
the centers of baryonic \cite{Ryan} and dark matter (DM)
\cite{DokEroPAZH} condensation into the growing protogalaxies.
There are proposed two alternative possibilities: (i) a formation
of initially massive primordial BHs and their successive growth up
to $\sim10^9{\rm M_{\odot}}$ due to accretion of ambient  matter
or (ii) a formation of small-mass primordial BHs and their
subsequent merging into the supermassive ones in the process of
hierarchical clustering of protogalaxies.

An effective cosmological mechanism of massive primordial BH
formation and their clusteri\-zation was developed in works
\cite{Ru1,Ru2,KR04}. In this papers the properties of spherically
symmetric primordial BH clusters were investigated. As a basic
example, a scalar field with the tilted Mexican hat potential had
been accepted. The properties of resulting primordial BH clusters
appear to be strongly dependent on the value of initial phase. In
addition, the properties of these clusters depend on the tilt
value of the potential $\Lambda $ and the scale of symmetry
breaking $f$ at the beginning of inflation stage. As a result, the
mass distribution of primordial BH clusters could vary in a wide
range. In our previous paper \cite{weqso} we considered the model
parameters leading initially to large clusters with a rather heavy
mass of the central primordial BH, $\sim4\,10^7{\rm M_{\odot}}$.
These central heavy primordial BH can grow due to accretion up to
$\sim10^{9}{\rm M_{\odot}}$ and, therefore, may explain the
observed early quasar activity.

The elaboration of a discussed mechanism of cosmological primordial
BH formation is far from completion and detailed elaboration. For
example, there are no now any physical substantiations for
reconstruction of scalar field potential parameters and initial
characteristics of primordial BH clusters. It is connected not only
with the uncertainties of observational data but also with
complexities of phase transition details. For example, the domain
walls formed during the phase transition in the early universe has a
topology of sphere but with a very complicated surface form. When
these closed domain walls are turned out inside the horizon, they
become self-gravitating. Inside the horizon domain walls tend to
obtain a spherical form due to surface tension, but at the same time
they strongly oscillate and generate gravitational and scalar waves.
As a result their mass gradually diminish. An approximate
consideration of this effect \cite{Ru2} demonstrate that for a wide
range of initial theoretical free parameters there are conditions
for formation of cluster with a supermassive primordial BH. For this
reason we will not fix here the definite values of free parameters
in the discussed cosmological model of massive black hole formation.
The influence of non-sphericity of the formed domain walls see in
\cite{Ru2,KR04}. An application of the mechanism found in
\cite{Ru1,Ru2,KR04} is not limited by a specific form of the scalar
field potential. Below we demonstrate that substantial number of
potentials, like e.~g. those used in hybrid inflation, also result
in formation of massive BHs. Moreover, it is hard to avoid
primordial BHs overproduction in the early Universe. In fact, any
inflationary model using potential with two or more minima must take
into account this mechanism of primordial BHs overproduction.

In this paper we choose parameters of the potential which lead to
formation of relatively small primordial BH clusters. We suppose
that relatively small primordial BH clusters provide the major
contribution to initial density pertur\-bations which afterwards
evolve into protogalaxies. The hierarchical clustering of
protogalaxies during the cosmological time leads to the observable
large scale structure. We describe the gravitational dynamics of DM
coupled with primordial BH clusters and demonstrate that a
protogalaxy could be formed without any initial fluctuations in DM
density. In this case the clusters of primordial BHs play the role
of initial fluctuations. Two scenarios of supermassive BHs formation
could coexist: (i) the most massive clusters of primordial BHs
account for an early quasar activity \cite{weqso}, but (ii) less
massive (considered in this paper) clusters of primordial BHs
produce more numerous supermassive BHs observed nowadays in almost
all structured galaxies.

There are several stages of BHs and galaxies formation in the
described scenario: (i) Formation of closed walls of scalar field
just after the end of inflation with a subsequent collapse some of
these walls with formation of massive primordial BH cluster with the
most massive BH in the center after the horizon crossing according
to \cite{Ru1,Ru2}. (ii) Detachment of the central dense region of
the primordial BH cluster from cosmological expansion and
virialization. Numerous small-mass BHs merge with a central one.
(iii) Detachment of the outer cluster region (where DM particles
dominate) from cosmological expansion and a protogalaxy growth.
Termination of a protogalaxy growth due to interaction with the
surrounding standard DM fluctuations. (iv) Gas cooling and star
formation accompanied by the merging of protogalaxies and final
formation of modern galaxies.

\section{Formation of primordial black holes in hybrid inflation}
\label{hybridsec}

\begin{figure}
 \label{hybridfig}
 \includegraphics[scale=0.2]{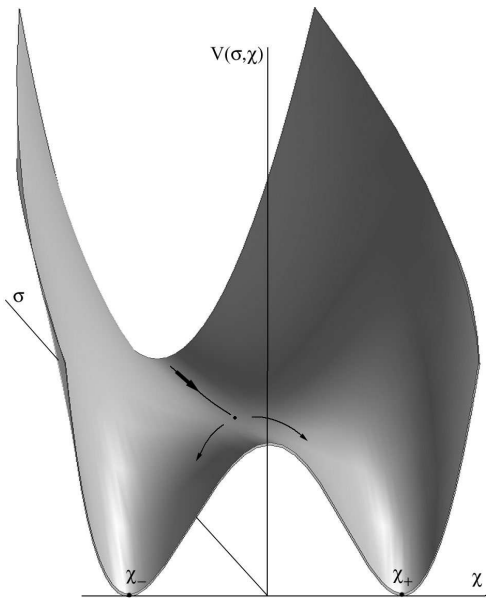}
 \caption{Potential for the hybrid inflation model. Arrows show the
 directions of classical motion of the evolving scalar field.}
\end{figure}

It is instructive to consider the mechanism of massive primordial BH
production in the framework of the hybrid inflationary model
\cite{Linde91b} following to results of paper \cite{Ru03a}.
According to \cite{Linde91b} the hybrid inflation potential has the
form
\begin{equation}
 \label{HibrPot}
 V(\chi,\sigma )=\varkappa ^{2}\left( M^{2}-\frac{\chi^{2}}{4}\right)
 ^{2}+\frac{ \lambda ^{2}}{4}\chi ^{2}\sigma
 ^{2}+\frac{1}{2}m^{2}\sigma ^{2}.
\end{equation}
Inflation proceeds during a slow rolling along the valley
$\chi=0,\sigma>\sigma _{c}$. When the field $\sigma$ decreases up
to $\sigma_{c}=\sqrt{2}\frac{\varkappa}{\lambda }M$, the motion
along the line $\chi =0$, $\sigma<\sigma _{c}$ becomes unstable,
and the field $\chi$ quickly moves to one of the minima $\chi
_{\pm }=\pm 2M,\sigma =0$, see Fig.~1. Inflation is finished
producing strong fluctuations around the accidentally chosen
minimum. This rather well elaborated picture suffers a serious
problem nevertheless. During an inflationary stage, when the
fields $\sigma$ and $\chi$ move classically along the line
$\chi=0$, the space is divided in many causally disconnected space
regions due to quantum fluctuations. The values of scalar fields
in neighboring regions (or domains) are slightly different. Each
e-fold time interval produces approximately $e^{3}\simeq20$
different regions. Hence there are about $e^{180}\simeq 10^{78}$
space domains right before the end of inflation. The values of
field in different domains chaotically distributed around the
point $\chi =0,\sigma =\sigma _{c}$. The domains with the field
value $\chi <0$ tend to the left minimum $\chi _{-}=- 2M,\sigma
=0$. Another part go to the right minimum $\chi _{+}=+ 2M,\sigma
=0$. A lot of walls between such domains appear and we run into
the well known problem of wall-dominated Universe
\cite{Zeldovich74}.

The only way for our Universe to evolve into recent state is to be
created with a nonzero initial field value, $\chi_{\rm in}\neq 0$ at
the beginning of inflation. During inflation, a nonzero field $\chi
$ is slowly approaching the critical line $\chi =0$. If, in the
middle of inflation, a field with average value approaches to the
critical line $\chi =0$, the fluctuations of the field in some part
of space domains could cross this line. In future, these domains
will be in a vacuum state, say, $\chi_{-}$ surrounded by a sea of
another vacuum $\chi_{+}$. The two vacua domains are separated by a
closed wall as it was discussed above. A number of these walls
strictly depends on the initial conditions at the moment of our
Universe creation, i.~e. at the beginning of inflation.

Let us estimate the energy and size of the formed closed walls by
supposing that a field in some space domain crosses the critical
line during the time corresponding to number $N$ of e-folds before
the end of inflation. A characteristic size of this domain is of the
order of the Hubble radius, $H^{-1}$, and it will increase
correspondingly in $e^{N}$ times up to the end of inflation. A
surface energy density of the domain wall after inflation for
potential \ref{HibrPot} is
\begin{equation}
 \epsilon =\frac{8\sqrt{2}}{3}\varkappa M^{3}.
 \label{sigma}
\end{equation}
A resulting total energy $E_{\rm wall}$ of the wall after inflation
is approximately
\begin{equation}
 E_{\rm wall}\simeq
 4\pi \epsilon \left( H^{-1}e^{N}\right)^{2}=
 4\sqrt{2}\frac{M_{\rm Pl}^{2}}{\varkappa M}e^{2N},
 \label{Ewall}
\end{equation}
where a numerical value of $N$\ is in interval $\left(
0<N<N_{U}\simeq 60\right)$. These walls collapse into BH with mass
$M_{\rm BH}\simeq E_{\rm wall}$ (see \cite{Ru2} for details). Let
us estimate the mass-scale of these BHs $M_{\rm BH}$ for the
characteristic values of parameters $ \varkappa =10^{-2}$ and
$M=10^{16}$~GeV. For $N=40$ we obtain $M_{\rm
BH}\simeq3\,10^{59}\mbox{~GeV}\sim 100{\rm M_{\odot}}$. The same
estimation of the minimum mass of BHs created at the e-fold number
$N=1$ before the end of inflation gives $M_{\rm
BH,min}\simeq10^{6}M_{\rm Pl}$. As a result the  hybrid inflation
leads to BH production with mass in the wide range
$10^{25}\hbox{~GeV}<M_{\rm BH}<10^2{\rm M_{\odot}}$. An abundance
of massive BHs depends on the proximity of an average field value
to critical the line $\chi =0$which in turn, depends on the
initial conditions and specific values of model parameters.

\begin{figure}
 \label{massprof}
 \includegraphics[angle=0,width=0.95\textwidth]{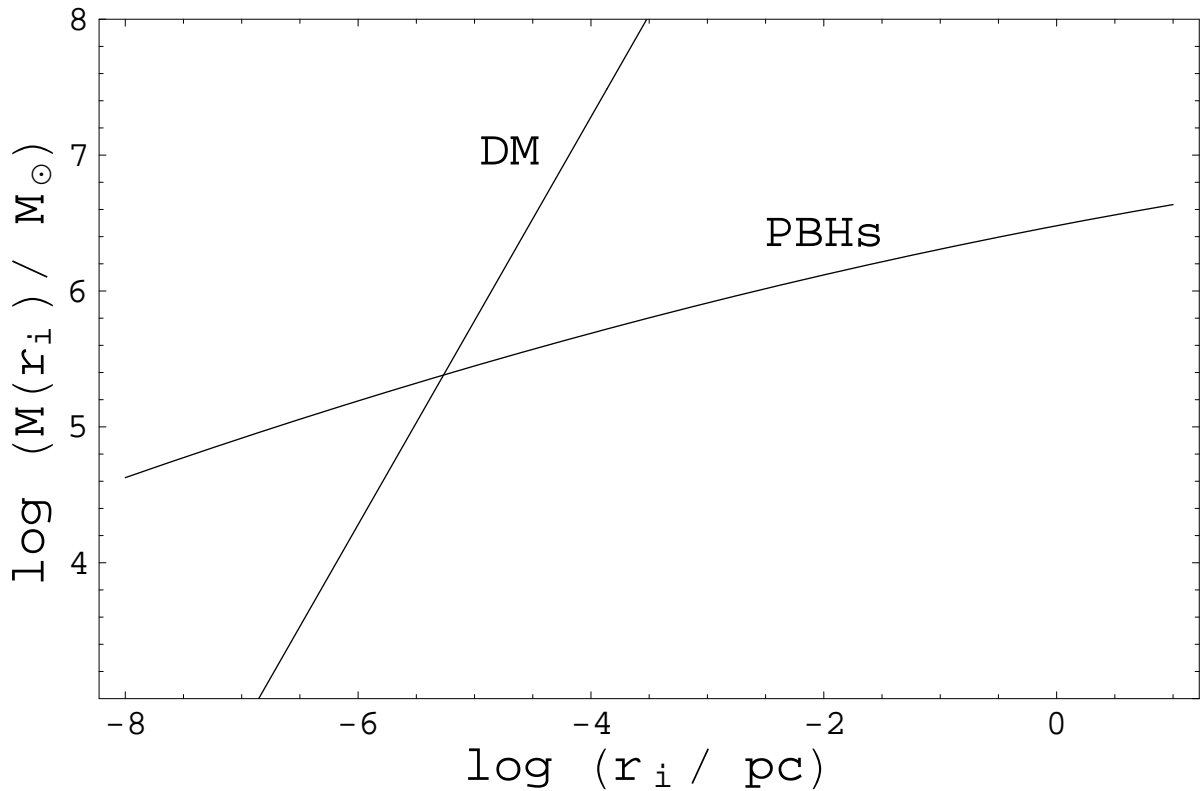}
 \caption{The initial mass profiles of primordial BHs $M_{h}(r_{i})$
 and DM $M_{\rm DM}(r_{i})$ in the cluster (protogalaxy).}
\end{figure}

The main finding of this consideration is that passive primordial BH
caused by phase transition during inflation is the rule rather than
exception. In this paper we elaborate the idea that clusters of
primordial BHs could be the seeds for galaxy formation. This work is
based on the results of \cite{Ru1,Ru2,KR04}, where the ``Mexican
hat'' potential was considered. Calculations based on this potential
provide the suitable framework for consideration of protogalaxies
formation around the clusters of primordial BHs.

An initial modeled mass profile $M_{h}(r_{i})$ of primordial BH in
the cluster (see details in \cite{Ru2a}) is shown in the Fig.~2.
This numerically calculated profile is a starting point for study
made below. For comparison in the Fig.~2 is shown also a radial
distribution of DM mass $M_{\rm DM}(r_{i})$ inside the same
sphere. Radius $r_{i}$ is a size of sphere at the moment $t_{i}$
and the temperature $T_{i}$, when this sphere is crossing the
cosmological horizon.

Note that different spheres in the Fig.~2 are shown at different
times $t_{i}$. Due to cosmological expansion the radial mass
distribution of uniform DM does not follow the law $M_{\rm
DM}\propto r^{3}$ as it must be for fixed time. A physical size of
chosen sphere at temperature $T_{i}$ is smaller than one in the
recent epoch in $T_{0}/T_{i}$ times, where $T_{0}=2.7$~K. A total
mass in the central parts of the BH cluster is so high, that some
part of BHs appear to be inside the combined gravitational radius
$r_g=2GM/c^2$. An initial total mass of BHs inside combined event
horizon is $2.7\,10^4M_{\odot}$. A massive BH of this mass becomes
the most massive central BH in the cluster.

\section{Gravitational dynamics of BH cluster and dark matter}
\label{dynamics}

Let us describe a gravitational dynamics of the primordial BH
cluster and internal DM in the combined gravitational field. In a
central part of the cluster the primordial BHs dominate in mass.
This central part of the cluster is detached from cosmological
expansion at the radiation dominated epoch. Conversely, the DM
dominates at the outer part of the cluster, and the outer part is
detached from cosmological expansion much more later at the matter
dominated epoch.

Consider a spherically symmetric system with radius $r<ct$,
consisting of (i) primordial BHs with a total mass $M_h$ inside the
radius $r$, (ii) radiation with energy density $\rho_r$, (iii) DM
with density $\rho_{\rm DM}$ and (iv) vacuum with an energy density
$\rho_{\Lambda}$. The radiation density (and obviously the density
of vacuum) is homogenous. The corresponding fluctuations induced by
primordial BHs are classified as entropy fluctuations. Because the
characteristic radial scale is much more less then the instant
horizon scale, we use in a standard way the Newtonian gravity but
will take into account the prescription of \cite{TolMc30} to treat
the gravity of homogenous relativistic components
$\rho\to\rho+3p/c^2$. The dynamical evolution of spherical shell
with an initial radius $r_i$ obeys the equation
\begin{equation}
 \frac{d^2r}{dt^2}=
 -\frac{G(M_h+M_{\rm DM})}{r^2}-\frac{8\pi G\rho_r r}{3}+
 \frac{8\pi G\rho_{\Lambda} r}{3},
 \label{d2rdt1}
\end{equation}
with an initial conditions at time $t_i$: $\dot r=Hr$ and
$r(t_i)=r_i $. Equation (\ref{d2rdt1}) is derived by taken into
account that $\varepsilon_r+3p_r=2\varepsilon_r$ and
$\varepsilon_{\Lambda}+3p_{\Lambda}=-2\varepsilon_{\Lambda}$. With
these initial conditions the shell initially is growing in radius
but expansion decelerates with time according to (\ref{d2rdt1}). At
some time the expansion is stopped, the shell separates from
cosmological expansion and start to shrink. All types of constituent
nonrelativistic internal matter in the cluster --- the dark matter,
primordial BHs and baryons follow the shell dynamics. As a result,
the solutions of (\ref{d2rdt1}) for shells with different initial
radii supply us with density distribution of the dark matter and
PBH. For numerical calculations it is useful to rewrite equation
(\ref{d2rdt1}) by using dimensionless variables:
\begin{equation}
 \label{bt}
 r(t)=\xi a(t)b(t),
\end{equation}
where $\xi$ is a comoving length, $a(t)$ is a dimensionless scale
factor of the universe normalized to the present moment $t_0$ as
$a(t_0)=1$ and dimensionless function $b(t)$ describes the
deflection of a chosen shell from the cosmological expansion (from
the Hubble law). A comoving length $\xi$ is related with a total
mass of DM inside considered spherical volume (i.~e. excluding
total BH mass) by the relation $M_{\rm DM}=(4\pi/3)\rho_{\rm
DM}(t_0)\xi^3$, where $\rho_{\rm DM}(t_0)$ is the nowadays DM
density. A scale factor $a(t)$ obeys one of the Friedman equation,
which can be rewritten as $\dot a/a=H_0E(z)$, where redshift
$z=a^{-1}-1$, $H_0$ is a present value of the Hubble constant and
function
\begin{equation}
 E(z)=
 [\Omega_{r,0}(1+z)^4+\Omega_{m,0}(1+z)^3+ \Omega_{\Lambda,0}]^{1/2},
 \label{efun}
\end{equation}
where $\Omega_{r,0}$ is the present density parameter of radiation,
$\Omega_{m,0}\simeq0.3$, $\Omega_{\Lambda,0}\simeq0.7$, and $h=0.7$.
By using the Friedman equation for $\ddot a$ one can rewrite an
evolution equation (\ref{d2rdt1}) as follows
\begin{equation}
 \frac{d^2b}{dz^2}+\frac{db}{dz}S(z)+\left(\frac{1+
 \delta_h}{b^2}-b\right)\frac{\Omega_{m,0}(1+z)}{2E^2(z)}=0,
 \label{d2bdz1}
\end{equation}
and $\delta_h=M_h/M_{\rm DM}$ is a fluctuation amplitude and
function
\begin{equation}
 S(z)=\frac{1}{E(z)}\frac{dE(z)}{dz}-\frac{1}{1+z}.
\end{equation}
In the limiting case $\Omega_{\Lambda}=0$ equation (\ref{d2bdz1}) is
equivalent to equation obtained in \cite{kt}. We start to trace out
the evolution of primordial BH cluster starting from an initial high
redshift $z_i$, when the considered shell crosses the cosmological
horizon $r\sim ct_i$. The initial conditions for this problem are
shown in the Fig.~\ref{massprof}.

The most early epoch in our calculation corresponds to formation
of the central most massive BH with a mass $2.7\,10^4M_{\odot}$,
described at the end of preceding Section. A corresponding
temperature of the universe at that time is $T\simeq16$~MeV. In
cosmological scenario with the standard perturbation spectrum a
mass of primordial BH could not exceed a total mass under the
instant horizon $M\sim (t/t_{\rm Pl})M_{\rm Pl}$ \cite{Carr94}. As
a result, at temperature $T\simeq16$~MeV the mass of primordial BH
cannot be larger than ~$10^3 M_{\odot}$. Nevertheless, in the
considered scenario the mass of primordial BHs is much more larger
because they are formed from the collapsing domain walls, not from
initial fluctuations. At the same time a total energy of domain
walls could be rather large because they are formed and stretched
during inflation. We suppose also that DM has been already
decoupled from radiation at this temperature. For example, in the
case neutralino DM particles with mass $100$~GeV and slepton mass
$1$~TeV a kinetic decoupling temperature is $\simeq150$~MeV
\cite{Schw03}, corresponding to a much more earlier epoch.
Therefore, the neutralino DM particles at the time of primordial
BH cluster formation are influenced only by gravitational forces
and the combine clustering of two-component medium (BHs$+$DM) is
described by a single equation (\ref{d2bdz1}) from the very
beginning. The same situation is realized for DM composed of
super-heavy particles with mass $m_{\chi}\sim10^{13}-10^{14}$~GeV
which probably never been in kinetic equilibrium with radiation.
In the opposite case (for some other DM particles candidates) the
growth of fluctuations in DM medium is suppressed by friction due
to interaction with radiation while BHs are clustering. The
super-heavy particles are more preferable for our model in
comparison with the neutralinos because their annihilation cross
section is very small $\propto m_{\chi}^{-2}$. In this case the
are no problems with a possible huge annihilation rate in the
central part of considered cluster.

\begin{figure}
 \label{massprof2}
 \includegraphics[angle=0,width=0.95\textwidth]{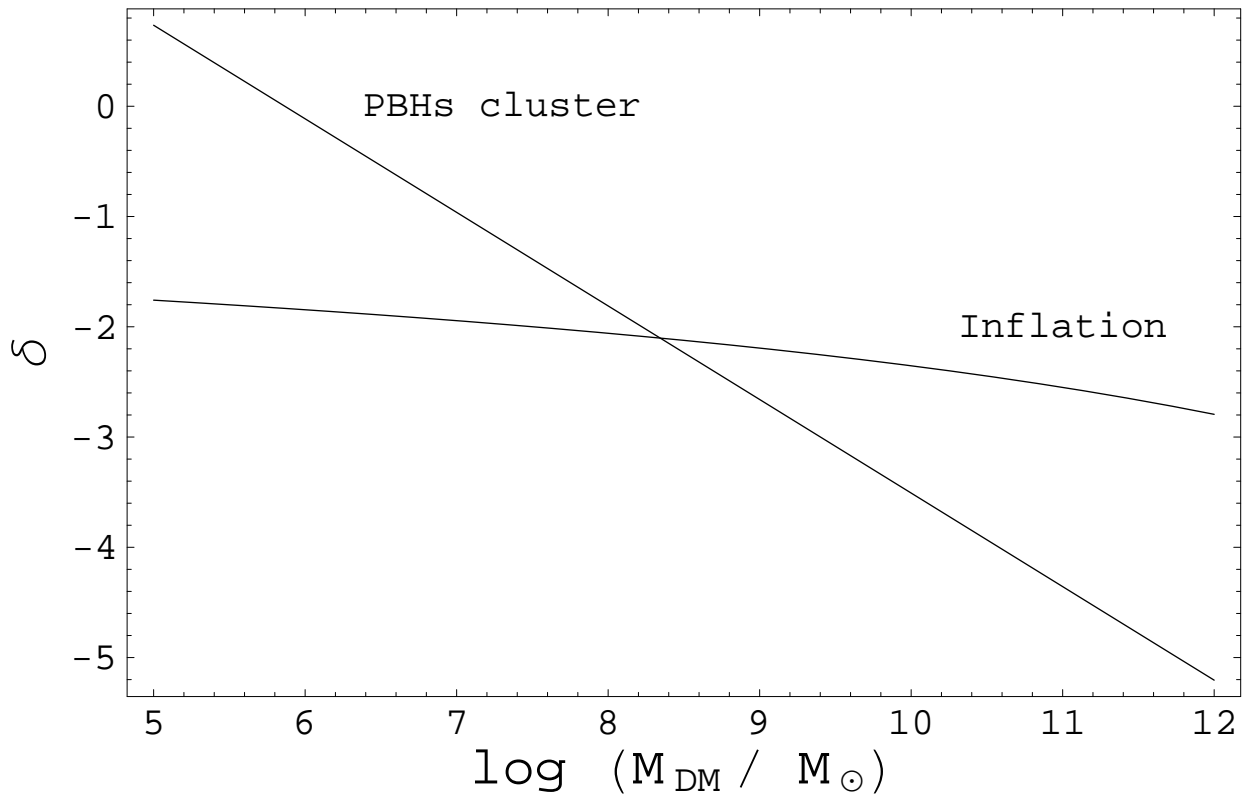}
 \caption{The r.m.s. density perturbations at the time $t_{\rm eq}$
 of matter-radiation equality are presented. Two cases are compared
 --- a total density fluctuations produced in the presence of
 primordial BH clusters and the standard ones generated by
 inflation.}
\end{figure}

The amplitude of fluctuations produced by primordial BH cluster
$\delta_{\rm eq}^{h}(M_{\rm DM})= 2.5\delta_{i}(M_{\rm DM})$ is
shown in the Fig.~3. The numerical factor $2.5$ corresponds to
entropy perturbations which grow up to time $t_{\rm eq}$ according
to the Meszaros solution \cite{peebles}. In the Fig.~3 the r.m.s.
values of standard inflation fluctuations are shown. For standard
DM fluctuations we use the fitting formula for power spectrum from
\cite{Dav85}:
\begin{equation}
 P(k)=\frac{Ak} {(1+1.71u+9u^{1.5}+u^2)^2},
 \label{cdm}
\end{equation}
where $u=k/[(\Omega_{m,0}+\Omega_{b,0})h^2\mbox{~Mpc}^{-1}]$, and
$k$ is a comoving wave vector in  Mpc$^{-1}$ units. The initial
spectrum is supposed to be the Harrison--Zeldovich type. The
relation between the mass scale $M$ of the r.m.s. perturbations and
the linear-scale $R$ is
\begin{equation}
\sigma(M)=\frac{1}{2\pi^2}\int\limits_0^\infty k^2\,dk\,P(k)W(k,R),
\label{sig}
\end{equation}
where $W(k,R)$ is a filtering function \cite{bardeen}. We put for
estimations $\delta_{\rm eq}\simeq \sigma_{\rm eq}$. The
normalization constant $A$ in (\ref{cdm}) corresponds to the
observable value $0.9$ of r.m.s. fluctuations at $8$~Mpc scale at
the recent time.

\begin{figure}
 \includegraphics[width=0.95\textwidth]{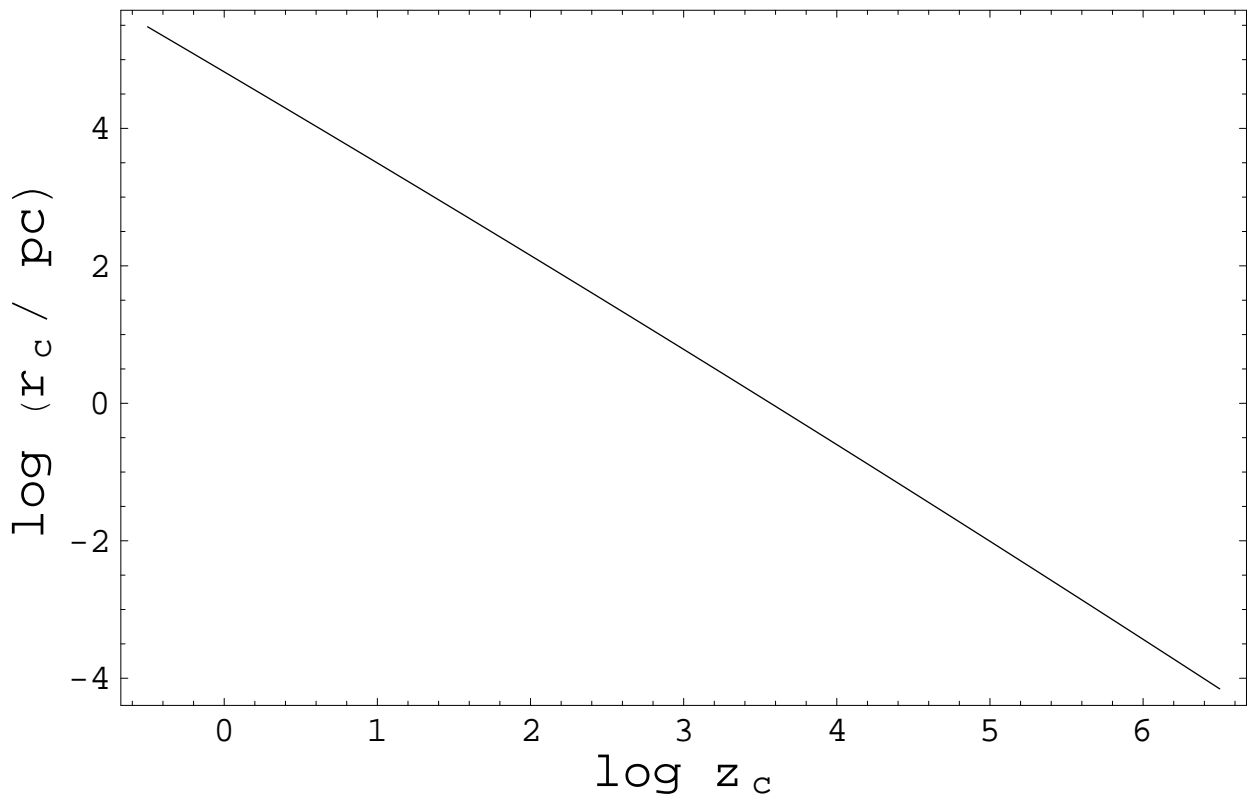}
 \caption{The virial radius of protogalaxy $r_{c}$ is shown as a
 function of redshift $z$.}
 \label{solgen}
\end{figure}

The termination of shell expansion, $\dot{r}=0$, at the instant of
time $t_s$ corresponds to condition $db/dz=b/(1+z)$ in accordance
with the definition of function $b$. Following to \cite{peebles} we
suppose that every chosen shell after the termination of expansion
is virialized and contracted from the maximum radius $r_{s}=r(t_s)$
to the radius $r_{c}=r_{s}/2$. The resulting average density of DM
in the virialized shell $\rho$ is 8 times larger than one at the
time of maximum shell expansion:
\begin{equation}
 \rho=8\rho_{m,0}(1+z_{s})^{3}b_{s}^{-3},
 \label{rhobs}
\end{equation}
where $b_s=b(t_s)$ and an effective (virialized) shell radius is
\begin{equation}
 r_{c}=\left(\frac{3}{4\pi}\frac{M_{\rm DM}}{\rho}\right)^{1/3}.
 \label{rcs}
\end{equation}
Numerical solution of equation (\ref{d2bdz1}) is shown in the
Fig.~\ref{solgen} and represents the growth of  protogalaxy radius
with time (or redshift $z$) in the absence of standard DM
fluctuations. This numerical solution is valid up to the time when
DM fluctuations start to grow effectively.

Let us trace a time evolution of the described spherical cluster
(protogalaxy) step by step starting from its central region. It is
obvious (and also validated by our numerical solution) that
expansion of more dense inner spherical shells stops earlier than
a corresponding expansion of rarefied outer shells. A central most
massive BH with mass $M_{c}=2.7\, 10^{4}M_{\odot}$ forms in the
the cluster of at the very early time as discussed in the
Sec.~{\ref{hybridsec}}. The dense central spherical shells are
detached from the cosmological expansion very early, at the
radiation dominated epoch. In in later time, nearly all matter of
these central shells will be accreted by the central BH  in the
process of two-body relaxation of BHs. We will describe this
process below.

The process similar to ``secondary accretion'' (i.~e. a
gravitational contraction of initially homogeneous DM around a
central mass) takes place for the early formed primordial BHs. As a
result, the cluster of primordial BHs would be ``enveloped'' by an
extended DM halo. We call these haloes the ``induced galaxies''
(IG). The resulting density profile in the cluster does not follow
the secondary accretion law $\rho\propto r^{-9/4}$ \cite{peebles}
because a central mass in our case is noncompact. The distribution
of DM in the cluster after the virialization is
\begin{equation}
 \rho_{\rm DM}(r)=\left.
 \frac{1}{4\pi r_c^{2}}\frac{dM_{\rm DM}(r_c)}{dr_c}\right|_{r_c=r},
 \label{dprofeq}
\end{equation}
where function $M_{\rm DM}(r_c)$ is determined from the solution of
equation (\ref{d2bdz1}), and $\rho$ and $r_c$ from the solution of
(\ref{rhobs}) and (\ref{rcs}) respectively.

\begin{figure}
 \includegraphics[width=0.95\textwidth]{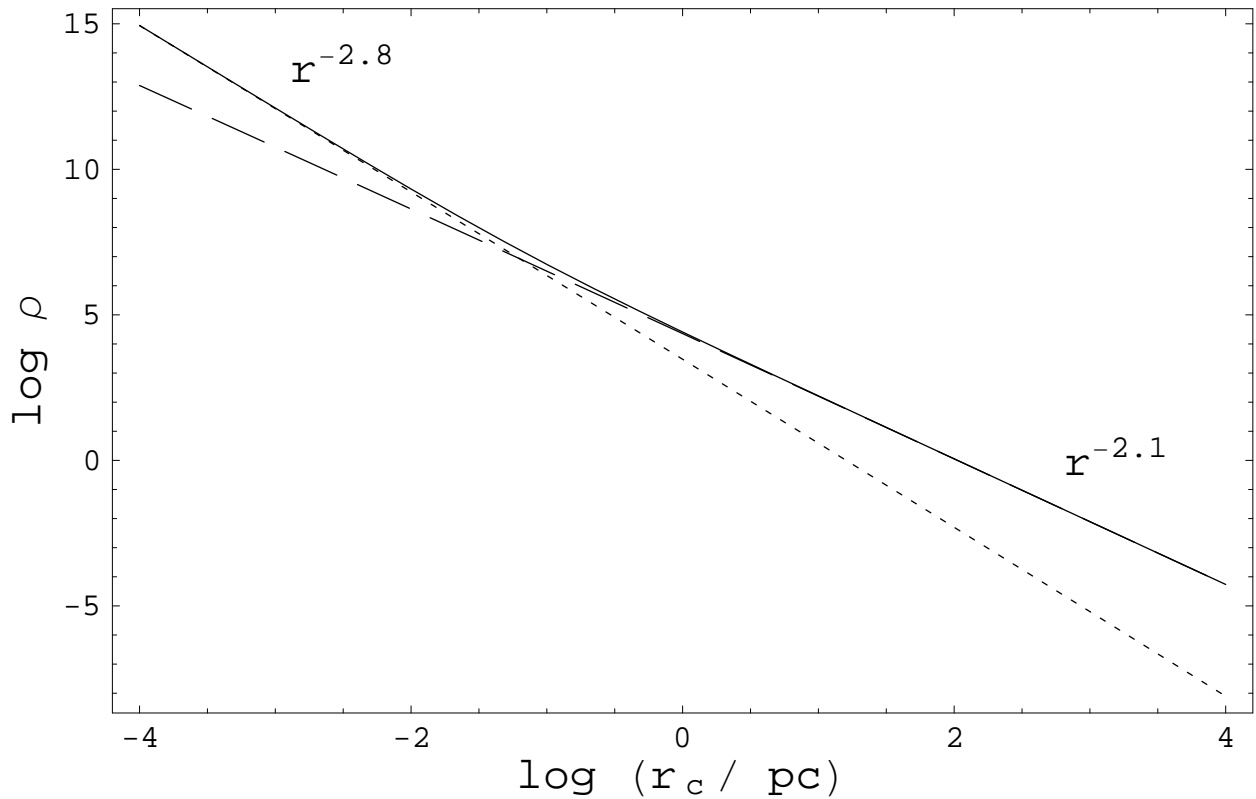}
 \caption{The final density profile of
 protogalaxy $\protect\rho(r_c)$ in units $M_{\odot}/{\rm pc}^{3}$ in
 dependence of radial distance $r_{c}$ from the cluster center for DM
 (dashed line), for BHs (dotted line) and for the sum of DM$+$BHs
 (solid line) respectively.}
 \label{solgen2}
\end{figure}

By analogy with a DM profile (\ref{dprofeq}) one can obtain the
corresponding profile for primordial BHs density $\rho_{\rm BH}(r)$
and for the total density $\rho_{\rm DM}(r)+\rho_{\rm BH}(r)$. The
results are shown in the Fig.~\ref{solgen2}, where density is
expressed in units $M_{\odot}/pc^{3}$ and radial distance is in
parsecs. These numerically calculated density profiles with a rather
good accuracy are fitted by power laws:
\begin{eqnarray}
 \label{dmfit1}
 \rho_{\rm DM}(r)&=&
 2.3\,10^4\left(\frac{r}{1\mbox{~pc}}\right)^{-2.13}
 {\rm M}_{\odot} \mbox{~pc}^{-3}, \\
 \rho_{\rm BH}(r)&=&
 2.9\,10^3\left(\frac{r}{1\mbox{~pc}}\right)^{-2.85}
 {\rm M}_{\odot}\mbox{~pc}^{-3}.
 \label{bhfit1}
\end{eqnarray}
At radial distance $r>0.056$~pc the local density of DM prevail over
the density of BHs, while a total internal mass of DM prevails over
a total mass of  BH at distance $r>0.7$~pc. Therefore, the influence
of BHs on a subsequent protogalaxy dynamics is limited by the
central parsec. This influence will be considered in the next
Section. The derived density profile (\ref{dmfit1}) differs from the
Navarro-Frenk-White or Moore et al. profiles, obtained in the
numerical simulations of DM halo formation but is very near to these
profiles at intermediate scales, when power-law index is $\simeq-2$.
An interesting properties of the derived density distribution is a
diminishing of mean virial velocity $V_v=(GM/2R)^{1/2}$ of IGs with
time (or with decreasing of $z$). This behavior is a consequence of
the specific shape of perturbation spectrum produced by clusters of
primordial BHs.

A total mass of IG is growing with time because increasingly more
distant regions are separated from cosmological expansion and
virialized around the central most massive BH. The growth of IG is
terminated at the epoch of a nonlinear growth of ambient standard
density fluctuations with a mass of the order of IG. These
fluctuations are originated in a standard way from inflation
cosmological perturbation spectrum $P(k)$ (see e.~g.
\cite{bardeen,mo02}. The fluctuations of both types are growing in a
similar way at the matter dominated epoch. Therefore, a
corresponding condition for termination of growth of typical IG due
to gravitational instability is
\begin{equation}
 \label{Deltaeq}
 \nu\sigma_{\rm eq}(M_{\rm DM})=\delta_{\rm eq}^h(M_{\rm DM}),
\end{equation}
where $\nu$ is the perturbations peak height (and we consider only
a mean perturba\-tion with $\nu=1$). The r.h.s of (\ref{Deltaeq})
is the of fluctuation caused by BH cluster. Respectively the
l.h.s. of (\ref{Deltaeq}) is the standard Gaussian fluctuations
according to (\ref{sig}). Both types of fluctuations are taken at
the moment of matter-radiation equality $t_{\rm eq}$. Numerical
solution of (\ref{Deltaeq}), which is the intersection point of
two curves in the Fig.~3, gives the final mass of IG
$M_{\mathrm{DM}}=2.2\,10^{8}M_{\odot}$.

The evolution of cosmological perturbations in the Universe with the
$\Lambda$--term at the matter dominated epoch can be derived from
equation (\ref{d2bdz1}) or from the corresponding equation of paper
\cite{mo02}:
\begin{eqnarray}
 \frac{\delta(t)}{\delta(z_{\rm eq})}\!=
 \!\frac{g(z)}{g(t_{\rm eq})}\frac{(1\!+\!z_{\rm eq})}{(1\!+\!z)},
 \quad g(z)\!\simeq\!\frac{5}{2}
 \frac{\Omega_m}{\Omega_m^{4/7}\!-\!\Omega_{\Lambda}\!+\!
 (1\!+\!\Omega_m/2)(1\!+\!\Omega_{\Lambda}/70)},
 \label{glam}
\end{eqnarray}
where $\Omega_m=\Omega_{m,0}(1+z)^3/E^2(z)$, $E(z)$ is from
(\ref{efun}) and $\Omega_{\Lambda}$ is defined in a similar way.
Now by fixing the perturbation amplitude $\delta(z_{\rm
eq})=\sigma_{\rm eq}(M_{\rm DM})$ at
$M_{\mathrm{DM}}=2.2\,10^{8}{\rm M_{\odot}}$, one can find from
(\ref{glam}) the instant of growth termination. This happens near
the redshift $z=15$, and so only in the narrow range of mass and
radius shown in the Figs.~2 and 4.

A resultant structure of IG is the following: The central BH with
mass $2.7\,10^4M_{\odot}$ is surrounded by the cluster of smaller
BHs with a total mass $2.2\,10^5M_{\odot}$ and with a radius
$r\sim0.7$~pc. Outside this sphere the DM prevail in mass and has
a density profile (\ref{dmfit1}). A total mass of mean IG (i.~e.
for $\nu=1$) is $2.2\,10^8M_{\odot}$ and the corresponding virial
radius of IG is $R=1.8$~kpc. In the inner parsec of the IG the
two-body relaxation and accretion processes operated from the very
outset. Starting from $z=15$ the IG participates in the
hierarchical clustering and a subsequent structure formation
proceeds by the standard scenario: small galaxies including IGs
are assembled into the larger galaxies, clusters and
superclusters. Being formed, the IG looks like a dwarf spheroidal
galaxy with a central massive BH surrounded by BHs of intermediate
mass and with a central DM density spike shown in the
Fig.~\ref{solgen2}. Some part of these IGs could escape galactic
merging and survive to our time.

\section{Accretion in the induced protogalaxy}
\label{intev}

Let us describe the accretion of DM and primordial BHs onto the
central BH in the described IG (or protogalaxy). The cluster of
primordial BHs is composed of BHs with different masses.
Therefore, an important factor of dynamical evolution is mass
segregation, i.~e. a concentration of more massive BHs closer to
the center. This mass segregation considerably complicates the
treatment of cluster dynamical evolution. We use here an
approximate approach by considering the BHs of different masses as
independent homological subsystems evolving in the combined
gravitational field. This approach is a similar one to used in
studies of evolution of multicomponent star clusters.

A total rate of DM accretion onto primordial BHs moving within a
radial distance $r$ from the IG center is
\begin{equation}
 \dot M_{\rm DM}=
 \sum_iN_i\sigma_{{\rm acc},i}v\rho_{\rm DM},
 \label{ac1}
\end{equation}
where $N_i$ is a total number of BHs with mass $M_i$ inside radius
$r$ (in reality the mass distribution of BHs is continuous),
$v\simeq(GM_{\rm tot}/r)^{1/2}$ is a mean (virial) velocity, $M_{\rm
tot}=M_{\rm DM}+\sum M_i N_i$, and a cross-section of DM particle
capture by BH is $\sigma_{{\rm acc},i}=\pi r_{g,i}(c/v)^2$,
$r_{g,i}=2GM_i/c^2$. A corresponding ``inverse'' characteristic
accretion time of DM is
\begin{equation}
 t_{\rm acc}^{-1}\simeq
 \frac{\dot M_{\rm DM}}{M_{\rm DM}}=\frac{3G^2}{cvr^3}\sum_iN_iM_i^2.
 \label{ac2}
\end{equation}
By using results of numerical solutions of the preceding Section we
find that accretion time of DM (\ref{ac2}) is a rather well
approximated by the power-law:
\begin{equation}
 t_{\rm acc}(r)\simeq
 8\,10^3\left(\frac{r}{1\mbox{~pc}}\right)^{2.7}\!\!t_0,
 \label{tac1}
\end{equation}
where $t_0$ is age of the Universe. From this relation it follows
that DM is totally accreted by now, $t_{\rm acc}\sim t_0$, inside
the radius $R_c\simeq0.036$~pc. Therefore, the DM density profile
(\ref{dmfit1}) is valid only at $r\geq R_c$. A total accreted mass
of DM is negligible in comparison with a total mass of BHs in the
cluster, and so the DM accretion is unimportant for the growth of
central BH.

A two-body relaxation time in the cluster of equal mass BHs is
\cite{Spit}:
\begin{equation}
 t_{\rm rel}\simeq\frac{1}{4\pi}\frac{v^{3}}{G^{2}m^{2}n\ln(0.4N)},
\label{trel}
\end{equation}
where $N$, $n$ and $m$ are respectively a total number, number
density and individual mass of BHs in the cluster. We will use now
for forthcoming estimations only mean values. A characteristic
time-life of BH cluster (due to ``evaporation'' of fast BHs) is
$t_{e} \simeq40t_{\rm rel}$ \cite{Spit}. At the end of this time the
gravitational collapse happens, starting with avalanche contraction
of the remaining BH cluster. The collapse proceeds shell by shell of
starting from the innermost shells. We estimate the corresponding
mean values of $t_{e}$ for shells with different radii by using
relation $t(z)=t_{e}$, where $t(z)$ is age of the Universe
corresponding to redshift $z$. As a result due to this dynamical
evolution process the collapsing shells of relaxed BHs increase the
mass of central BH.

As well as accretion of DM, the accretion of collapsing shells of
primordial BHs provides a rather small contribution to the growth of
central BH in the IG.

Indeed, a mass of central BH at $z=15$, when a growth of IG is
terminated, is $M_{\mathrm{BH} }=6.9\,10^{4}M_{\odot}$. This mass
is the sum of initial central BH mass $2.7\,10^{4}M_{\odot}$ and a
total mass of collapsed shells of primordial BHs at that epoch. At
the time of large galaxy formation, corresponding to $z\simeq1.7$,
the central BH mass is $M_{\mathrm{BH} }=7.2\,10^{4}M_{\odot}$. If
some IG survived up to nowadays epoch, $z=0$, it will have now the
central BH with mass $M_{\mathrm{BH}}=7.3\,10^{4}M_{\odot}$ (due
to accretion of DM and primordial BHs). From these estimations we
conclude that a main contribution to the growth of central
supermassive BHs in galaxies was provided not by the DM and
primordial BH accretion but accretion of baryonic matter (gas
and/or stars) and/or merging of galaxies.

\section{Merging of protogalaxies and black holes}
\label{merge}

In Section~\ref{dynamics} we calculated a characteristic mass of
IG or protogalaxy $2.2\,10^{8}M_{\odot}$ which is formed around a
seed primordial BH cluster at $z=15$. At this epoch in the
vicinity of considered IG there are a lot of smaller neighboring
protogalaxies, both ordinary and IG. All these protogalaxies will
hierarchically merge later into the large modern galaxies.

The individual IGs are massive enough to sink down by dynamical
friction into the galactic center during the Hubble time. The mass
loss of IGs due to tidal stripping in spiralling down to the
galactic is ineffective due to their large density. Indeed, a
condition for tidal stripping of particles at distance $r_s$ from IG
center and at distance $r$ from the host galaxy center is the
equality of acceleration produced by IG and the tidal acceleration:
\begin{equation}
  \frac{GM(r_s)}{r_s^2}=r_s\frac{d}{dr}\frac{GM_{\rm H}(r)}{r^2},
 \label{tid1}
\end{equation}
where $M(r_s)$ and $M_{\rm H}(r)$ are respectively the mass profile
of IG and the host galaxy. According to the Navarro-Frenk-White
model the DM distribution in the Galactic halo is
\begin{equation}
 \rho_{\rm H}(r)=
 \frac{\rho_{0}}{\left(r/L\right)\left(1+r/L\right)^2},
 \label{halonfw}
\end{equation}
where $L=28$~kpc, $\rho_{0}=5\,10^6M_{\odot}$kpc$^{-3}$ and for
normalization it used the local density in the Sun vicinity. Using
the density profile (\ref{dmfit1}) we find from (\ref{tid1}) that
at any radial distance $r$ the radius of tidal stripping $r_s$ is
greater than IG radius $1.8$~kpc. Therefore, the IGs sink down to
the galactic center as a whole without tidal stripping.

The possibility of IG spiralling down to the galactic center by the
influence of dynamical friction depends on the initial orbit of IG.
Suppose for estimation that orbit of IG is circular. By using the
known expression for dynamical friction force \cite{Sasl} and an
equation for the angular momentum loss, one finds the differential
equation for orbital radius evolution
\begin{equation}
 \frac{dr}{dt}=
 -\frac{4\pi G^2M_s(r)\rho_{\rm H}(r)\Lambda Br}{v(r)^3},
 \label{difdyn}
\end{equation}
where $v(r)=\sqrt{GM_H(r)/r}$, $\Lambda\simeq10$, $B\simeq0.427$.
As it was shown above, the tidal stripping is ineffective and so
the mass of IG $M_s(r)=const=2.2\,10^8M_{\odot}$. Consider at
first the density profile (\ref{halonfw}) for our Galaxy. It was
formed when age of the Universe was approximately one fourth of
the nowadays days age. Numerical solution of equation
(\ref{difdyn}) demonstrates that only IGs inside the radius
$26$~kpc have enough time to sink down into the Galactic center.
The typical elliptical galaxies are formed earlier than our Galaxy
and being much more denser. Therefore, all IGs in elliptical
galaxies sank down to their centers.

According to observations the masses of central supermassive BHs in
Sa, Sb, Sc galaxies are in general smaller than those in E and S0
galaxies. In our model this is connected with a relatively late
formation of Sa, Sb, Sc galaxies, when the main part of primordial
BHs have not enough time to sink down to the galactic center. In
particular, $\sim10^3$ BHs with mass $\sim10^5M_{\odot}$ enveloped
by IG can inhabit in our Galaxy. They could be observed as the
widely discussed ultra-luminous X-ray sources.

The fate of primordial BHs inside the central parsec of the host
galaxy is rather uncertain. We suppose that during the Hubble time
major part of these BHs are merged into a single supermassive BH.
Namely, the dynamical friction is a very effective mechanism for BH
merging at final stage because the density of IG, $\rho\propto
r^{-2.8}$ (see in the Fig.~\ref{solgen2}), is strongly growing
towards the center and smoothed out only at very small distance
$R_c\simeq0.036$~pc from the central BH. An additional dynamical
force is produced by interactions of IGs with stars from the bulge
and central star cluster. As a result, the late phase of BHs merging
proceeds very fast. The probability of simultaneous presence in the
galactic nucleus of three or more BHs is very low due to the
slingshot effect. On the contrary, a substantial amount of massive
BHs may inhabit the galactic halo, if they turn out rather far from
the galactic center from the very beginning \cite{Ru1,Ru2,KR04}. Our
assumption of multiple merging of primordial BHs may be violated in
the less dense galaxies of late Hubble types.

Multiple coalescence of massive primordial BHs in the galaxies is
inevitably accompanied by the strong burst of gravitational
radiation. The future interfero\-metric detector LISA is capable to
detect these coalescence events. A simple estimation of the burst
rate from an observable part of the Universe gives
\begin {equation}
 \dot N_{grav}\sim\frac{4\pi}{3}\frac{N}{t_0}(ct_0)^3~n_g \sim100
 \left(\frac{n_g}{10^{-2}\mbox{Mpc$^{-3}$}}\right)
 \left(\frac{t_0}{10^{10}\mbox{yrs}}\right)^2
 \left(\frac{N}{10^3}\right)\mbox{~yrs$^{-1}$},
\end{equation}
where $n_g$ is a mean number density of structured galaxies
(galaxies with nuclei) and $N$ is a mean number of merging events
per galaxy. Gravitational bursts provide the principal possibility
for the verification of considered model by the LISA detector.

\section{Correlations of central black holes with bulges}

Recent observations (see e.~g. \cite{Gebh}) reveal the correlations
between the mass of the central Supermassive BH (SBH) in the
galactic nucleus $M_{\rm SBH}$ and velocity dispersion $\sigma_e$ at
the bulge half-optical-radius:
\begin{equation}
 M_{\rm BH}=1.2(\pm0.2)\,10^8\left(\frac{\sigma_e}{200\mbox{~km/s}}
 \right)^{3.75(\pm0.3)}{\rm M}_{\odot}.
 \label{korsig}
\end{equation}
In other set of observations \cite{Ferra} a different form of
correlation was derived: $M_{\rm SBH}\propto\sigma_e^{4.8(\pm0.5)}$.
We show below that our model of IGs reproduces correlations
(\ref{korsig}). At the early stage of hierarchical clustering of
small protogalaxies into the bigger ones the discussed primordial
black holes are homogenously mixed with DM at the scales greater
than IGs. For this reason a total mass of these primordial BHs in
any galaxy $\sum M_{\rm BH}$ would be proportional to the galactic
DM halo mass $M$. After the final merging of primordial BHs into a
single central BH the similar relation retains: $M_{\rm
SBH}\propto\sum M_{\rm BH}\propto M$. By taking in mind that
velocity dispersion in galaxy is determined mainly by DM, one may
expect the existence of some relation between $M_{\rm SBH}$ and
$\sigma_e$. We find the form of this relation in the following way.
The condition for galaxy formation from a density fluctuation
$\delta$ is
\begin{equation}
 \label{forcond}
 \delta_c=
 \delta_{\rm eq}(M)\frac{g(z)(1+z_{\rm eq})}{g(z_{\rm eq})(1+z)},
\end{equation}
where the function $g(z)$ is from (\ref{glam}), $\delta_c=1.686$ is
the threshold value of fluctuation for spherical collapse and $M$ is
a mass of the galaxy. For fluctuation amplitude $\delta_{\rm eq}(M)$
we take the r.m.s. fluctuation (\ref{sig}), and so we neglect the
distributions but take into account only the mean values. Equation
(\ref{forcond}) gives implicitly the functional dependance $z(M)$.
The density of a virialized object in $\varkappa=18\pi^2$ times
greater than the mean cosmological density
$\rho_m(z)=\rho_{c,0}\Omega_{m,0}(1+z)^3$ of DM at the time
corresponding to redshift $z$. This provides us with the relations
between the radius, velocity dispersion in formed galaxy and
galactic mass:
\begin{equation}
 \label{radden}
 r(M,z)=\left[\frac{3M}{4\pi\varkappa\rho_m(z)}\right]^{1/3}, \quad
 \sigma_e(M)=\left[\frac{GM}{r(M,z(M))}\right]^{1/2},
\end{equation}
where $z(M)$ is derived from (\ref{forcond}). By inverting the
function $\sigma_e(M)$ and by using DM fluctuation spectrum
(\ref{cdm}) we find numerically
\begin{equation}
 \label{mv200}
 M\simeq7\,10^{11}
 \left(\frac{\sigma_e}{200\mbox{~km~s}^{-1}}\right)^{4.3}{\rm M}_{\odot}.
\end{equation}
If merging of primordial BHs into the one supermassive central BH
proceeds effectively, from this relation we find a resulting mass of
the central supermassive BH:
\begin{equation}
 M_{\rm BH}=\psi\Omega_hM=1.4\,10^8\left(\frac{\psi\Omega_{h}}
 {2\,10^{-4}}\right)
 \left(\frac{\sigma_e}{200\mbox{~km~s}^{-1}}\right)^{4.3},
 \label{bhn}
\end{equation}
where a factor $\psi$ is related with a possible additional growth
of the central BH by accretion of DM and baryonic matter. This model
is in a reasonable agreement with observation data (\ref{korsig}).
The derived power index $\alpha$ in relation $M_{\rm
SBH}\propto\sigma_e^\alpha$ is closer to one obtained in
\cite{Ferra}. This power index in our model is completely defined by
fluctuation spectrum at the galactic scales or, more definitely, by
the value $n\simeq-2$ of power index. A possible dependence of an
additional factor $\psi$ on the mass, $M$, could modify a functional
relation $M_{\rm SBH}=M_{\rm SBH}(\sigma_e)$. Nevertheless, a simple
case $\psi=const$ provides a good agreement of the derived $M_{\rm
SBH}(\sigma_e)$ relation with observations (\ref{korsig}). This
relation is naturally realized in the model without accretion
$\psi=1$. We expect that a minor influence of accretion or universal
accretion fraction $\psi=const$ in the resulting mass of the central
supermassive BHs in the galactic centers may be explained in
detailed gas dynamics models of galactic nuclei.

It must be noted that $M_{\rm SBH}-\sigma_e$ correlation is a
general feature of stochastic mechanism of supermassive BHs
formation and is revealed also in other models of primordial BHs
formation, e.~g. \cite{DokEroPAZH}.

\section{Discussion}

We describe here a new model of protogalaxy formation with the
cluster of primordial BHs as a source of initial density
perturbation. The used mechanism of primordial BH formation
\cite{KR04,Ru2a} provide us with a set of primordial BH clusters of
different total mass. This variety of initial conditions leads,
therefore, to the variety of protogalaxies from the very beginning
of their formation. In this paper we choose for numerical modeling
only those BH clusters which produce the large number of small
relatively protogalaxies. This model predicts the very early galaxy
and quasar formation. An other inevitable consequence of this model
it the existence of intermediate mass BHs beyond the dynamical
centers of galaxies and in the intergalactic medium. May be one of
these type intermediate mass BHs was already observed by the X-ray
Chandra telescope in the galaxy M82 \cite{Kaar00}.

More definitely in this model the protogalaxies are formed at
redshift $z=15$. These induced protogalaxies have initially the
following parameters: a constituent total mass of DM $M_{\rm
DM}=2.2\,10^{8}{\rm M_{\odot}}$, a virial radius $1.8$~kpc, a mass
of central BH $M_{\mathrm{BH}}=7.2\,10^{4}M_{\odot}$. In the
following cosmological and dynamical evolution, these
protogalaxies are assembled by hierarchical clustering into the
nowadays galaxies. The clustering process occurs in a stochastic
manner and leads to the specific correlation between the central
supermassive BH mass and galactic bulge velocity dispersion
\cite{DokEroPAZH}. An alternative proposed scenario is based on
the initial large primordial BH clusters, when a resulting galaxy
contains a single primordial BH growing due to accretion of
ambient gas and stars and producing early quasar activity
\cite{weqso}.

It is worth to estimate in the framework of our model a
probability to find a nowadays galaxy without supermassive BH.
Induced galaxies (with a central cluster of primordial BHs) and
ordinary small protogalaxies have mass
$M_{\mathrm{DM}}=10^{8}M_{\odot}$, while the modern galaxies are
much more massive, $M_{\rm DM}=10^{12}{\rm M}_{\odot}$. The
merging of induced galaxy with an ordinary protogalaxy produces a
next generation protogalaxy with the massive central BH.
Therefore, about $10^{4}$ collisions is required to form the
nowadays galaxy. Suppose that an amount of induced galaxies is
about 0.1\% comparing with the ordinary ones. A corresponding
probability to find a modern galaxy without supermassive BH is
less than $0.999^{10000}\simeq 4.5\,10^{-5}$. Hence, even a very
small fraction of induced galaxies is able to explain the
observable abundance of AGN.

\begin{acknowledgments}
The work of V.I.D. and Yu.N.E. been supported in part by the Russian
Foundation for Basic Research grants 06-02-16029 and 06-02-16342,
the Russian Ministry of Science grants LSS 4407.2006.2 and LSS
5573.2006.2.
\end{acknowledgments}

\end{document}